\documentclass{emulateapj} 
 
\newcommand{\lsim}{\raisebox{-0.13cm}{~\shortstack{$<$ \\[-0.07cm] $\sim$}}~}
\newcommand{\gsim}{\raisebox{-0.13cm}{~\shortstack{$>$ \\[-0.07cm] $\sim$}}~}
\newcommand{\tfm}{$\rm 24 \, \mu m \,$}

\shorttitle{The close environment of \tfm galaxies at $0.6<\lowercase{z}<1.0$}
\shortauthors{K. I. Caputi et al.}

\begin{document} 
 
 
%________________________________________________________________ 
\title{The close environment of \tfm galaxies at $0.6<\lowercase{z}<1.0$ in the COSMOS field
}

 
%________________________________________________________________ 
\author{K. I. \ Caputi\altaffilmark{1,2}, 
K. Kova\v{c}\altaffilmark{1},
M. Bolzonella\altaffilmark{3},
S. J. Lilly\altaffilmark{1},
G. Zamorani\altaffilmark{3},
H.  Aussel\altaffilmark{4},
D. Sanders\altaffilmark{5},
S. Bardelli\altaffilmark{3},
A. Bongiorno\altaffilmark{6},
T. Contini\altaffilmark{7},
G. Coppa\altaffilmark{3},
O. Cucciati\altaffilmark{8},
S. de la Torre\altaffilmark{9},
L. de Ravel\altaffilmark{9},
P. Franzetti\altaffilmark{10},
D. Frayer\altaffilmark{11}, 
B. Garilli\altaffilmark{10},
A. Iovino\altaffilmark{7},
P. Kampczyk\altaffilmark{1},
J.-P. Kneib\altaffilmark{9},
C. Knobel\altaffilmark{1},
F. Lamareille\altaffilmark{7},
J. F. Le Borgne\altaffilmark{7},
V. Le Brun\altaffilmark{9},
O. Le F\`evre\altaffilmark{9},
E. Le Floc'h\altaffilmark{5},
A. Leauthaud\altaffilmark{9},
C. Maier\altaffilmark{1}, 
V. Mainieri\altaffilmark{12},
M. Mignoli\altaffilmark{3}, 
R. Pell\`o\altaffilmark{7},
Y. Peng\altaffilmark{1},
E. P\'erez-Montero\altaffilmark{7},
E. Ricciardelli\altaffilmark{13},
M. Salvato\altaffilmark{14},
M. Scodeggio\altaffilmark{10},
N. Scoville\altaffilmark{14},
J. Silverman\altaffilmark{1}, 
J. Surace\altaffilmark{11},
M. Tanaka\altaffilmark{12},
L. Tasca\altaffilmark{9},  
L. Tresse\altaffilmark{9},
D. Vergani\altaffilmark{10},
E. Zucca\altaffilmark{3},
U. Abbas\altaffilmark{9},
D. Bottini\altaffilmark{10},
P. Capak\altaffilmark{14}, 
A. Cappi\altaffilmark{3},
C. M. Carollo\altaffilmark{1},
P. Cassata\altaffilmark{9},
A. Cimatti\altaffilmark{15},
M. Fumana\altaffilmark{10},
O. Ilbert\altaffilmark{5},
J. Kartaltepe\altaffilmark{5},
D. Maccagni\altaffilmark{10},
C. Marinoni\altaffilmark{9},
H. McCracken\altaffilmark{16},
P. Memeo\altaffilmark{10},
B. Meneux\altaffilmark{6},
P. Oesch \altaffilmark{1},
C. Porciani\altaffilmark{1},
L. Pozzetti\altaffilmark{3},
A. Renzini\altaffilmark{13},
R. Scaramella\altaffilmark{17},
C. Scarlata\altaffilmark{14}
 }

\altaffiltext{1}{Institute of Astronomy, Swiss Federal Institute of Technology (ETH H\"onggerberg), CH-8093, Z\"urich, Switzerland.}
\altaffiltext{2}{E-mail address: caputi@phys.ethz.ch}
\altaffiltext{3}{INAF Osservatorio Astronomico di Bologna, Bologna, Italy.}
\altaffiltext{4}{CEA/DSM-CNRS, Universit\'e Paris Diderot, DAPNIA/SAp, Orme des Merisiers, 91191 Gif-sur-Yvette, France.}
\altaffiltext{5}{Institute for Astronomy, University of Hawaii, Honololu, HI, USA.}
\altaffiltext{6}{Max Planck Institut f\"ur Extraterrestrische Physik, Garching, Germany.}
\altaffiltext{7}{Laboratoire d'Astrophysique de Toulouse-Tarbes, Universit\'e de Toulouse, CNRS, France.}
\altaffiltext{8}{INAF Osservatorio Astronomico di Brera, Milano, Italy}
\altaffiltext{9}{Laboratoire d'Astrophysique de Marseille, France.}
\altaffiltext{10}{INAF-IASF Milano, Milan, Italy.}
\altaffiltext{11}{Spitzer Science Center. California Institute of Technology, Pasadena, CA, USA.}
\altaffiltext{12}{European Southern Observatory, Garching, Germany.}
\altaffiltext{13}{Dipartimento di Astronomia, Universit\`a di Padova, Padova, Italy.}
\altaffiltext{14}{California Institute of Technology, Pasadena, CA, USA.}
\altaffiltext{15}{Dipartamento di Astronomia, Universit\`a degli Studi di Bologna, Italy}
\altaffiltext{16}{Institut d'Astrophysique de Paris, UMR 7095 CNRS, Universit\'e Pierre et Marie Curie, 98 bis Boulevard Arago, F-75014 Paris, France.}
\altaffiltext{17}{INAF, Osservatorio di Roma, Monteporzio Catone (RM), Italy}



\begin{abstract}

We investigate the close environment of 203 {\em Spitzer} \tfm-selected sources at $0.6<z<1.0$ using zCOSMOS-bright redshifts and spectra of $I<22.5$ AB mag galaxies, over 1.5 deg$^2$ of the COSMOS field.   We quantify the degree of passivity of the LIRG and ULIRG environments by analysing the fraction of close neighbours with $\rm D_n \,(4000)>1.4$. We find that LIRGs at $0.6<z<0.8$ live in more passive environments than those of other optical galaxies that have the same stellar mass distribution. Instead, ULIRGs inhabit more active regions (e.g. LIRGs and ULIRGs at $0.6<z<0.8$ have, respectively, $(42.0 \pm 4.9)\%$ and $(24.5 \pm 5.9)\%$ of  neighbours with $\rm D_n \,(4000)>1.4$ within 1~Mpc and $\pm \, 500 \, \rm km \, s^{-1}$). The contrast between the activities of the close environments of LIRGs and ULIRGs appears  especially enhanced in the COSMOS field density peak at $z\sim0.67$, because LIRGs on this peak  have a larger fraction of passive neighbours, while ULIRGs have as active close environments  as those outside the large-scale structure.  The differential environmental activity is related to the differences in the distributions of stellar mass ratios between LIRGs/ULIRGs and their close neighbours, as well as in the general local density fields. At $0.8<z<1.0$, instead, we find no differences in the environment densities of ULIRGs and other similarly massive galaxies, in spite of the differential activities. We discuss a possible scenario to explain these findings.

\end{abstract}



%%%%%%%%%%%%%%%%%%%%%%%%%%%%%%%%%%%%%%%%%%%%%%%%%%%%%%%%%%%%%%%%%%%%%%%%%%%
\section{Introduction}
\label{sec-intro}

 Our knowledge of the nature of the sources composing the extragalactic IR background (Puget et al.~1996; Dole et al.~2006) has been much clarified over the last years thanks to studies conducted with the {\em Spitzer Space Telescope} (Werner et al.~2004) and now also the {\em Akari Telescope} (Matsuhara et al.~2006). Many properties of these galaxies are now quite well known, including their redshift distribution (e.g. Rowan-Robinson et al.~2005; Caputi et al.~2006a; Desai et al.~2008); luminosity evolution (e.g. Le Floc'h et al.~2005; P\'erez-Gonz\'alez et al.~2005; Babbedge et al.~2006; Caputi et al.~2007);   assembled stellar masses (Daddi et al.~2005; Caputi et al.~2006a,b; Papovich et al.~2006; Noeske et al.~2007), and multiwavelength spectral characteristics (e.g. Houck et al.~2005; Yan et al.~2005; Choi et al.~2006; Weedman et al.~2006; Berta et al.~2007; Lacy et al.~2007; Sajina et al.~2007; Caputi et al.~2008; Rigby et al.~2008).

 The properties of the environment in which IR-selected galaxies reside are less well-determined. This is  mainly because environmental studies require the combination of wide area surveys with accurate redshift determinations  that, in general, are only provided  by large spectroscopic samples. Recent works have explored the link between environment and star-formation activity in IR-selected galaxies. Elbaz et al.~(2007) determined that the star-formation rate (SFR)-density relation observed locally was reversed at $z\sim1$. Serjeant et al.~(2008) suggested that this reversed relation could also hold for sub-millimetre galaxies at $1.0<z<1.5$.  Marcillac et al. (2008) found that luminous and ultra-luminous IR galaxies (LIRGs and ULIRGs, respectively) at $0.7<z<1.0$ are present, on average, in denser environments than those of other optical galaxies, but they argued that this difference is mainly driven by stellar mass.

The Cosmic Evolution Survey (COSMOS; Scoville et al.~2007)  has been designed to probe galaxy evolution and the effects of environment up to high redshifts over 2 deg$^2$ of the sky.  COSMOS observations include the coverage of the field with the Multiband Photometer for {\em Spitzer} (MIPS; Rieke et al.~2004) as part of two  {\em Spitzer} Legacy Programs (S-COSMOS; Sanders et al. 2007). COSMOS also comprises a large spectroscopic follow-up (zCOSMOS; Lilly et al.~2007) with the  Visible Multiobject Spectrograph (VIMOS; Le F\`evre et al.~2003) on the {\em Very Large Telescope (VLT)}.  The combination of the S-COSMOS and zCOSMOS datasets are ideal to characterise the environment of IR-selected galaxies and investigate the properties of their close neighbours. 

In this paper, we make use of zCOSMOS-bright spectra to  investigate the characteristics of the environment of the most luminous \tfm-selected galaxies with $I<22.5$ AB mag at redshifts $0.6<z<1.0$.  Firstly, in \S\ref{sec_closeenv}, we analyse the activity and stellar masses of galaxies which are close to \tfm galaxies of different IR luminosities. Secondly, in \S\ref{sec_denspeak},  we particularly study the activity of close neighbours of IR galaxies lying on the COSMOS field density peak at $z\sim0.67$.  Finally, in \S\ref{sec_densf}, we characterise the environment of \tfm sources using the general density fields produced with zCOSMOS (Kova\v{c} et al.~2008). We adopt throughout a cosmology with $\rm H_0=70 \,{\rm km \, s^{-1} Mpc^{-1}}$, $\rm \Omega_M=0.3$ and $\rm \Omega_\Lambda=0.7$.

%%%%%%%%%%%%%%%%%%%%%%%%%%%%%%%%%%%%%%%%%%%%%%%%%%%%%%%%%%%%%%%%%%%%%%%%%%%%%%%%%%%%%%%%%%%%%%%%%%%%%%%%%
\section{Sample selection and close environment definition}
\label{sec_sample}

The zCOSMOS-bright survey is now half-way complete and has already produced reduced spectra and redshifts for 10,644 sources over 1.5 deg$^2$ of the COSMOS field (`the 10k sample' hereafter). Most of the targets for zCOSMOS-bright are selected randomly from a complete $I<22.5$ AB mag catalogue (the `parent catalogue').  The sampling rate of the parent catalogue achieved with the 10k sample is  $\sim 1/3$. All zCOSMOS-bright observations are performed with the $R \sim 600$ VIMOS MR grism, with a spectral coverage of $(5500-9500) \, \rm \AA$ and a dispersion of 2.55 $\rm \AA$ (Lilly et al.~2007).

Using the zCOSMOS-bright 10k sample, Caputi et al.~(2008) identified 609 {\em Spitzer}/MIPS \tfm-selected galaxies with $S_{24 \, \rm \mu m}> 0.30 \, \rm mJy$ and $I<22.5$ AB mag, over 1.5 deg$^2$ of the COSMOS field. The number of identified galaxies is the result of the optical $I<22.5$ AB mag selection and the zCOSMOS-bright 10k sampling rate. As it was explained by Caputi et al.~(2008), $66\%$ of the  $S_{24 \, \rm \mu m}> 0.30 \, \rm mJy$ sources have an association in the zCOSMOS-bright parent catalogue and $\gsim 80\%$ of the missing sources are at $z>1$ (as determined from photometric redshifts; Ilbert et al.~2008; P. Oesch et~al., in preparation). Thus, the $I<22.5$ AB mag cut is sufficient for the identification of the vast majority of the IR-bright galaxy population at $z<1.0$.

Caputi et al.~(2008) analysed different optical spectral properties for their IR galaxies, depending on redshift. In particular, 217 out of the 609  \tfm galaxies have been identified to be at redshifts $0.6<z<1.0$. In this redshift range, the 4000~$\rm \AA$ break is included among the features observed in the most secure part of the zCOSMOS  spectra (i.e. far from the borders of the spectral coverage and, particularly, outside the region affected by fringing $\lambda \gsim 8500 \, \rm \AA$). The strength of the $4000 \,\rm \AA$ break $\rm D_n$(4000) is directly related to the time since the last episode of star formation, so this parameter is very useful to quantify galaxy activity.

  We use the zCOSMOS-bright 10k sample to identify and characterise the $I<22.5$ AB mag galaxies that live in the close environment of IR galaxies at $0.6<z<1.0$.   As in Caputi et al.~(2008), we only consider sources with reliable spectroscopic redshifts,  as defined by their zCOSMOS flags (see Lilly et al.~2007). 
  
\begin{figure}[!ht]
\begin{center}
\epsscale{.6}
\plotone{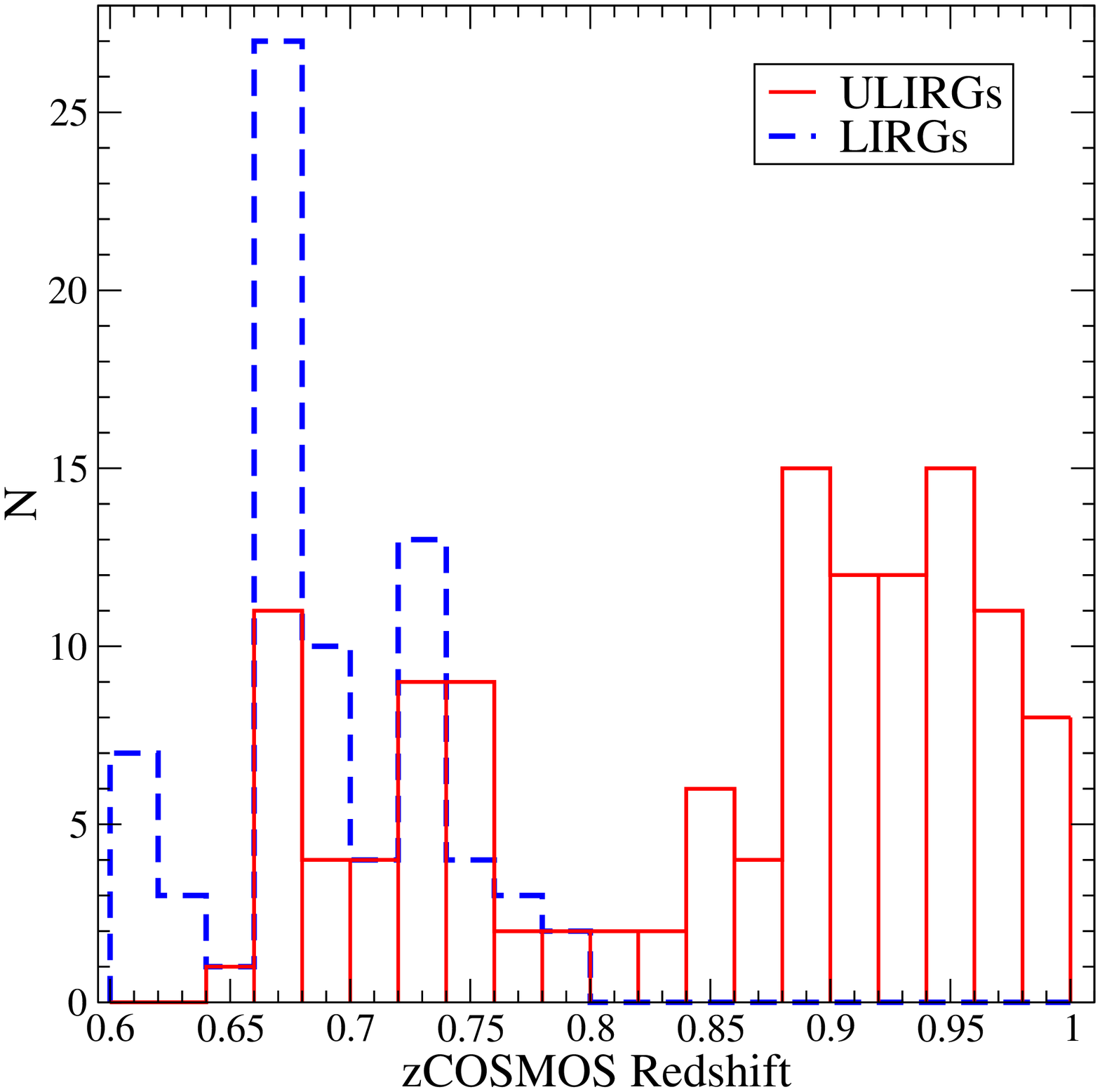}
\end{center}
\caption[]{\label{fig_zhisto} The redshift distribution of the 203 \tfm-selected galaxies considered in this work. The terms LIRGs and ULIRGs refer to galaxies with rest-frame \tfm luminosities below and above $\nu L_\nu^{\rm 24 \, \mu m} = 1.5 \times 10^{11}\, \rm L_{\odot}$, respectively. This cut has been chosen for being the luminosity completeness limit in the $0.8<z<1.0$ redshift bin. Note that the flux limit  $S_{24 \, \rm \mu m}> 0.30 \, \rm mJy$ of the S-COSMOS shallow survey also imposes a completeness limit for the LIRG population observed at $0.6<z<0.8$, which is $\nu L_\nu^{\rm 24 \, \mu m} = 7 \times 10^{10}\, \rm L_{\odot}$ (equivalent to a bolometric IR luminosity $L_{\rm bol.}\approx 3.4 \times 10^{11} \, \rm L_{\odot}$, using the Bavouzet et al.~2008 calibration).}
\end{figure}

  We analyse separately the neighbours of sources with rest-frame IR luminosities $\nu L_\nu^{\rm 24 \, \mu m} < 1.5 \times 10^{11}\, \rm  L_{\odot} \,$  and $\nu L_\nu^{\rm 24 \, \mu m} > 1.5 \times 10^{11}\, \rm  L_{\odot}$, as only the latter IR luminosities are above the completeness limits of the $S_{24 \, \rm \mu m}> 0.30 \, \rm mJy$ survey at $0.8<z<1.0$.  The redshift distribution of the 203 considered \tfm-selected galaxies is shown in figure \ref{fig_zhisto}. There are additional 14 IR sources in the Caputi et al.~(2008) sample at $0.8<z<1.0$, but their IR luminosities are below the completeness limit for this redshift bin. These 14 sources are statistically insufficient to reach any significant conclusion on the $\nu L_\nu^{\rm 24 \, \mu m} < 1.5 \times 10^{11}\, \rm  L_{\odot} \,$ population at $0.8<z<1.0$, so we do not consider them in our analysis.

   Following the most recent calibration for the conversion between monochromatic  ($\nu L_\nu^{\rm 24 \, \mu m}$) and bolometric ($L_{\rm bol.}$) IR luminosities made by Bavouzet et al.~(2008), we obtain that $\nu L_\nu^{\rm 24 \, \mu m} = 1.5 \times 10^{11}\, \rm L_{\odot}$ corresponds to $L_{\rm bol.}\approx 6 \times 10^{11} \, \rm L_{\odot}$.  For the same rest-frame \tfm luminosity, other previous calibrations would yield:  $L_{\rm bol.}\approx 7 \times 10^{11} \, \rm L_{\odot}$ (using the $\nu L_\nu^{\rm 25 \, \mu m}$-$L_{\rm bol.}$ relation by Takeuchi et al.~2005) and $L_{\rm bol.}\approx 1.3 \times 10^{12} \, \rm L_{\odot}$ (Sajina et al.~2006). Thus, our IR luminosity cut $\nu L_\nu^{\rm 24 \, \mu m} = 1.5 \times 10^{11}\, \rm L_{\odot}$  very roughly separates LIRGs and ULIRGs, respectively (whose nominal separation is $L_{\rm bol.} = 10^{12} \, \rm L_{\odot}$). In the following, we use the names LIRGs and ULIRGs  to refer to the IR sources below and above our luminosity completeness limit in the highest redshift bin, which is  $\nu L_\nu^{\rm 24 \, \mu m} = 1.5 \times 10^{11}\, \rm L_{\odot}$.

 The flux cut of the S-COSMOS $S_{24 \, \rm \mu m}> 0.30 \, \rm mJy$ survey also imposes a lower limit on the IR luminosities of the LIRGs we detect at $0.6<z<0.8$. We estimate that our completeness limit in this redshift bin is   $\nu L_\nu^{\rm 24 \, \mu m} \approx 7 \times 10^{10}\, \rm L_{\odot}$, which corresponds to $L_{\rm bol.}\approx 3.4 \times 10^{11} \, \rm L_{\odot}$ (Bavouzet et al.~2008).

  To select the neighbours in the close environment of the zCOSMOS-identified \tfm galaxies at $0.6<z<1.0$, we look for the sources in the zCOSMOS-bright 10k catalogue that lie within cylinders centred at the zCOSMOS position and redshift (RA, DEC, $z$) of each  \tfm source. We consider cylinders with a projected physical radius\footnote{$r=D_A \, \Delta \theta$, where $D_A$ is the angular diameter distance in Mpc and $\Delta \theta$ is the angular separation in radians.} of up to 2 Mpc and a velocity dispersion of $\rm \pm 500 \, km \, s^{-1}$. This velocity dispersion --which is $\sim 5$ times the typical uncertainty of the zCOSMOS-bright redshifts-- is small enough to expect the physical association of the neighbours with the central \tfm source. Also, such small dispersions ensure that the neighbours are observed while witnessing the IR phase of the central galaxy, which typically lasts $10^7$-$10^8$ years (Caputi et al.~2008).

 The total numbers of non-repeated neighbours  within 2 Mpc are 220 and 131, for 74 LIRGs and 42 ULIRGs at  $0.6<z<0.8$, respectively, and 154 for 87 ULIRGs at $0.8<z<1.0$.  We note that only $\sim 6\%$ of these neighbours  are also \tfm galaxies with fluxes $S_{24 \, \rm \mu m}> 0.30 \, \rm mJy$.
  
 We define three control samples, containing optical galaxies selected to mimic the stellar mass distributions of LIRGs and ULIRGs at $0.6<z<0.8$, and ULIRGs at $0.8<z<1.0$, respectively.  By using this set of mass-selected control samples for each specific population, we can probe whether the differences in the environmental properties of IR and other optical galaxies are driven or not by differences in stellar mass.

  To construct control samples that mimic the stellar mass distributions of each IR galaxy population, we select those (non-IR) zCOSMOS-bright galaxies whose stellar masses are within $\pm$0.2 dex of the stellar mass of each IR source. To guarantee that the resulting stellar mass distributions are the same as for IR galaxies, we weigh each galaxy in the control sample by a factor of $f=(n^i_{IR} \, N_c) / (n^i_c \, N_{IR})$, where $N_{IR}$ and $N_c$ are the total numbers of galaxies in the IR and control samples, respectively, and  $n^i_{IR}$ and $n^i_c$ are the respective numbers of galaxies in each stellar mass bin.

%%%%%%%%%%%%%%%%%%%%%%%%%%%%%%%%%%%%%%%%%%%%%%%%%%%%%%%%%%%%%%%%%%%%%%%%%%%%%%%%%%%%%%%%
\section{The close environment of \tfm galaxies at $0.6<\lowercase{z}<1.0$}
\label{sec_closeenv}

\subsection{The activity of close \tfm-galaxy neighbours}
\label{sec_activ}

We study the fraction  of active galaxies around  LIRGs and ULIRGs at $0.6<z<0.8$, and ULIRGs at $0.8<z<1.0$, as a function of their projected physical distance. We quantify the galaxy activity through the strength of the 4000$\rm \AA$ break $\rm D_n(4000)$, defined as the ratio of the average flux density between the (4000-4100) and (3850-3950)$\rm \AA$ bands (Balogh et al.~1999). The parameter $\rm D_n(4000)$ has the advantage of being quite independent of reddening (e.g. a typical extinction value $A_V \approx 1$ only produces a correction of $\sim 4\%$), so we do not apply any extinction correction to our measured $\rm D_n(4000)$ values.

Caputi et al.~(2008) showed that all bright \tfm-selected galaxies at redshifts $0.6<z<1.0$ are characterised by small values of the $\rm D_n(4000)$ parameter, typically $\rm D_n(4000)\lsim 1.2$. They interpreted these small values as the consequence of  rejuvenating bursts of  star formation which are also responsible for the IR activity of these galaxies. On the other hand, large values of the  $\rm D_n(4000)$ parameter such as $\rm D_n(4000)\gsim 1.4$ correspond to galaxies which have not suffered any new burst of star formation for at least $\sim 0.5$ Gyr (see Caputi et al. 2008). Thus, by measuring the fraction of galaxies with large  $\rm D_n(4000)$ values,  one can quantify the degree of `passivity' (or inactivity) of a given galaxy population.

We note that a large $\rm D_n(4000)$ parameter value does not exclude the possibility of minor star-formation activity, but indicates that the level of on-going star formation is insignificant with respect to the galaxy already-assembled stellar mass.  Also, in principle, very obscured ($A_{\rm V}>4-5$) bursts of star formation can as well produce large $\rm D_n(4000)$ values, due to differential extinction. However, such obscured bursts are very rare among optically bright galaxies. In particular, none of our galaxy neighbours in the zCOSMOS-bright 10k sample at $0.6<z<1.0$ has a best-fit UV through near-IR spectral energy distribution (SED) with $A_{\rm V}>4$ (Bolzonella et al.~2008).

The $\rm D_n(4000)$ parameter value has a non-negligible dependence on galaxy metallicity for ages $\gsim 1 \, \rm Gyr$ (Kauffmann et al.~2003). However, we note that by imposing a cut 
$\rm D_n(4000)\gsim 1.4$, we only select galaxies whose last burst of star formation happened 
at least $\sim 0.5$ Gyr before, independently of their metallicity.

\begin{figure*}[!ht]
\begin{center}
\plotone{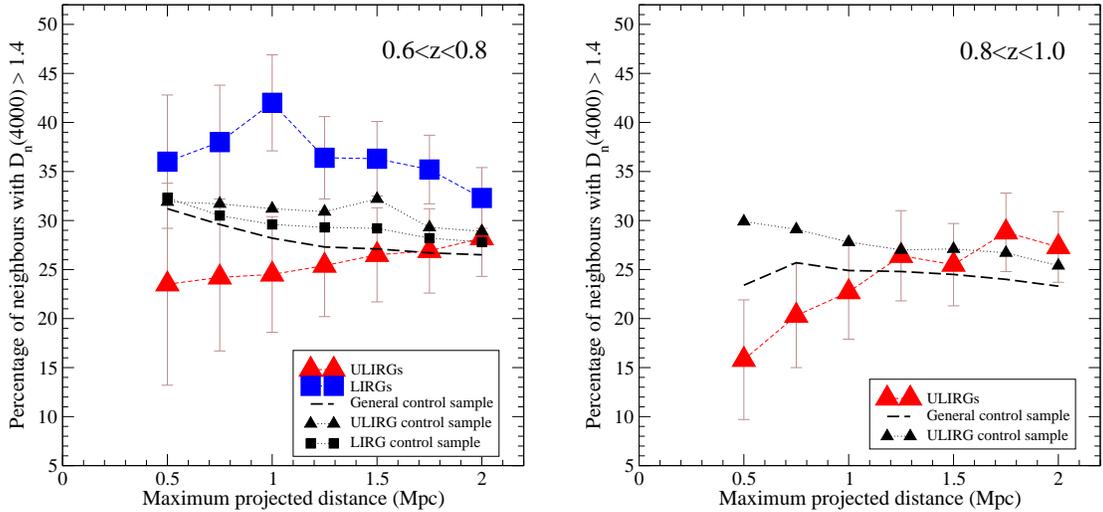}
\end{center}
\caption[]{\label{fig_dn4000} Percentages of (cumulative) neighbours with $\rm D_n(4000)> 1.4$ for LIRGs (large blue squares), ULIRGs (large red triangles) and for galaxies in their respective control samples (small black squares and triangles), versus the maximum projected physical distance. All neighbours are selected within $\pm 500 \, \rm km \, s^{-1}$ of the central source.  The dashed lines indicate the fraction of neighbours with $\rm D_n(4000)> 1.4$ for  generic control samples that include all non-IR zCOSMOS galaxies, independently of their stellar mass.}
\end{figure*}

Figure \ref{fig_dn4000} shows the percentages of (cumulative) close neighbours of \tfm galaxies which have $\rm D_n(4000)> 1.4$, versus the maximum projected physical distance, at $0.6<z<0.8$ (left-hand panel) and  $0.8<z<1.0$ (right-hand panel).   As a reference, we also show the fraction of close neighbours with $\rm D_n(4000) > 1.4$ for galaxies in the control samples (small black symbols) and for galaxies in a generic control sample that includes all non-IR optical galaxies with no mass selection (dashed lines).  For the mass-selected control samples, each neighbour has been assigned the weight factor of its central galaxy, or the average of weight factors when it is associated with more than one galaxy.

We compute the error bars on each percentage by assuming a binomial distribution, where each neighbour galaxy has two possibilities, either having $\rm D_n(4000)> 1.4$ or $< 1.4$. The error bars for the control samples are negligible in comparison to those for IR galaxies. The errors on the individual $\rm D_n(4000)$ measurements  are only random and $<8\%$ in all cases.

The plot in the left-hand panel of Figure \ref{fig_dn4000} shows that, at $0.6<z<0.8$, the fraction of close passive ($\rm D_n(4000) > 1.4$) neighbours is significantly different for LIRGs and ULIRGs. The maximum difference is observed within a projected physical distance of 1 Mpc, where the percentage of passive neighbours of ULIRGs is only $(24.5 \pm 5.9)\%$, while the percentage for LIRGs is $(42.0 \pm 4.9)\%$.

This difference is not an effect of stellar mass. When we compare the fractions of passive neighbours of LIRGs and ULIRGs with those of other similarly massive galaxies, we still see that LIRGs prefer more passive environments, while ULIRGs prefer more active ones. At $0.8<z<1.0$, ULIRGs also live in more active environments than those of other similarly massive galaxies, but this effect is only seen within less than $1 \, \rm Mpc$.

When considering up to larger scales, the differences in the fractions of passive neighbours are smaller. Within $\sim$2 Mpc of projected physical distance,  the fractions of passive sources around LIRGs, ULIRGs and other optical galaxies are all very similar. Besides, our results show that these fractions within 2~Mpc very little evolve from $0.8<z<1.0$ to $0.6<z<0.8$.

 If we adopted a more strict cut for the $\rm D_n(4000)$ parameter value, we would also observe a differential activity  for the LIRG and ULIRG neighbours at $0.6<z<0.8$.  At a maximum projected physical distance of 1 Mpc, where the differences are the largest, ULIRGs have $(18.9\pm 5.3)\%$ of neighbours with $\rm D_n(4000) > 1.5$, while LIRGs have $(33.0\pm 4.7)\%$. When considering a $\rm D_n(4000) > 1.6$ cut, the percentages of passive neighbours are  $(13.2\pm 4.7)\%$ and $(25.0\pm 4.3)\%$, for ULIRGs and LIRGs, respectively. At $0.8<z<1.0$, ULIRGs have similar fractions of $\rm D_n(4000) > 1.5$ or 1.6 neighbours as ULIRGs at $0.6<z<0.8$.

Of course, our selection  of close neighbours within a velocity dispersion of $\pm 500 \, \rm km \, s^{-1}$ from the central source, and the measurement of the $\rm D_n(4000)$ parameter, require the availability of spectra. Thus, it is not possible for us to conclude on the presence and activity of other possible close neighbours rather than those sampled by the zCOSMOS-bright 10k sample. However, it is worth noting that most of the results presented in this work correspond to fractions rather than absolute numbers, so the effects of incompleteness in  the sampling of $I<22.5$ AB mag galaxies should have very little impact (if any) on the quoted values.

In any case, to show that there is no statistical bias in the sampling of neighbours of LIRGs and ULIRGs, we perform the following test. We use the very good quality photometric redshifts ($z_{\rm phot}$) obtained in the COSMOS field (Ilbert et al.~2008), to select all those $I<22.5$ AB mag sources that lie within a projected physical distance of 2 Mpc and $\Delta z= \pm 1\sigma$ of each IR source, where $\sigma$ is the typical dispersion of the COSMOS photometric redshifts. For $I<22.5$ AB mag sources at $z<1$, Ilbert et al. obtain $\sigma=0.007 \times (1+z)$ (note that this corresponds to a velocity dispersion more than four times larger than the one we use to define the close environment in this work). Out of these $I<22.5$ AB mag $z_{phot}$-defined neighbours, we check the fractions of those included in the zCOSMOS-bright 10k sample, for our different IR-galaxy populations. We find that 28.0 and 29.0\% of the  $z_{\rm phot}$-defined neighbours around our $0.6<z<0.8$ LIRGs and ULIRGs, respectively, are included in the zCOSMOS-bright 10k sample. For the respective control samples, the percentages are very similar (28.7 and 29.1\%). At $0.8<z<1.0$, we find that 31.4\% of the ULIRG $z_{\rm phot}$-neighbours have a spectroscopic redshift, while the control sample neighbours are targeted at a rate of 29.1\%. The similarity of these percentages shows that there is no statistical bias in the spectroscopic sampling of neighbours for LIRGs and ULIRGs, or their respective control samples. These percentages are consistent with the overall sampling rate of $\sim 1/3$ estimated for $I<22.5$  AB mag galaxies in the zCOSMOS-bright 10k sample.

\subsection{The stellar masses of close \tfm-galaxy neighbours}
\label{sec_envstm}

\begin{figure}
\epsscale{.7}
\begin{center}
\plotone{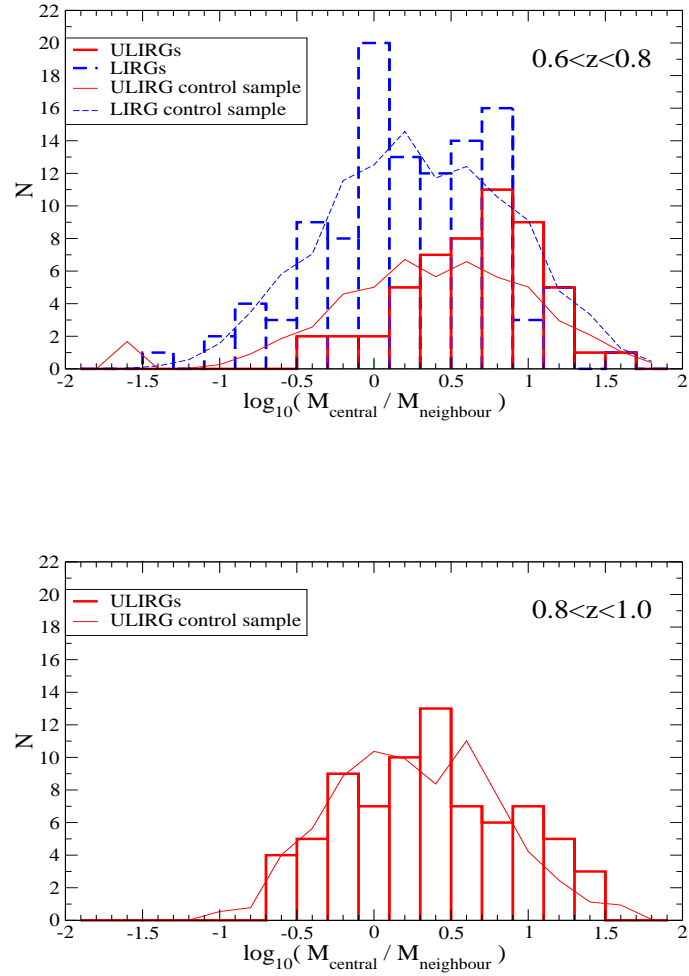}
\end{center}
\caption[]{\label{fig_mass} Histograms of stellar mass ratios between  \tfm galaxies and each of their close zCOSMOS neighbours within 1 Mpc of projected physical distance. The thin lines show the re-normalised distributions for galaxies in the respective control samples.}
\end{figure}

To investigate whether the differential activity of the close environments of LIRGs and ULIRGs is related to other environmental differences, we study the distribution of stellar mass ratios ($\rm M_{central}/M_{neighbour}$) between the central source and each of its neighbours within 1~Mpc of projected physical distance (Figure \ref{fig_mass}).  The red solid and blue dashed histograms in Figure \ref{fig_mass} correspond to ULIRG-neighbour and LIRG-neighbour pairs, respectively. The thin lines with the same style code show the re-normalised numbers of central-neighbour pairs for galaxies in the corresponding control samples.

All stellar masses have been taken from  Bolzonella et al.~(2008), who obtained UV through $4.5 \, \rm \mu m$ best-fit SEDs and derived stellar masses for all galaxies in the zCOSMOS-bright 10k sample. The $\chi^2$ best-fit SEDs have been determined using HYPERZ (Bolzonella et al.~2000) and the Bruzual and Charlot (2003; also Bruzual 2007) template library with solar metallicity. For the fitting of each source SED, the redshift has been fixed to the known spectroscopic value, while the star-formation history, age and reddening have been taken as free parameters. All stellar masses considered here correspond to a Chabrier (2003) initial mass function (IMF).

The two histograms in the top panel of Figure \ref{fig_mass}  show that the stellar mass ratios  ($\rm M_{central}/M_{neighbour}$) are substantially different for LIRGs and ULIRGs. While $\sim 30$\% of  LIRGs have smaller stellar masses than their close neighbours, ULIRGs are the most massive galaxy of their surroundings in more than 90\% of the cases.

Comparison with the respective control samples shows that there are also some differences in the ($\rm M_{central}/M_{neighbour}$) distributions  for IR galaxies and other optical galaxies with equivalent stellar masses at $0.6<z<0.8$. We perform a Kolmogorov-Smirnov test for each IR galaxy population and its respective control sample, properly incorporating the weight factors for the latter. We find that the null hypothesis that the ($\rm M_{central}/M_{neighbour}$) distributions for ULIRGs and their control sample at $0.6<z<0.8$ are drawn from the same parent sample can be rejected at a 99.9\% level (i.e. within our sample, ULIRGs are surrounded by less massive neighbours than other galaxies with similar stellar masses). For LIRGs and their control sample at $0.6<z<0.8$, the rejection of the null hypothesis is weaker, but still can be done at a 76\% level.  For ULIRGs at $0.8<z<1.0$, the null hypothesis cannot be rejected (56\% confidence for the rejection).

%%%%%%%%%%%%%%%%%%%%%%%%%%%%%%%%%%%%%%%%%%%%%%%%%%%%%%%%%%%%%%%%%%%%%%%%%%%%%%%%%%%%%%%%%%%%
\section{The close environment of \tfm galaxies in the density peak at $\lowercase{z}\sim0.67$}
\label{sec_denspeak}

Figure \ref{fig_zhisto} shows that 36.5\% of our LIRGs and 26.2\% of our ULIRGs at $0.6<z<0.8$  lie on a density peak at $0.66<z<0.68$. It is then interesting to compare the activity of the IR-galaxy close neighbours appearing in this narrow redshift range, with the global neighbour activity analysed in Section \ref{sec_activ}.

\begin{figure}[!ht]
\begin{center}
\epsscale{.8}
\plotone{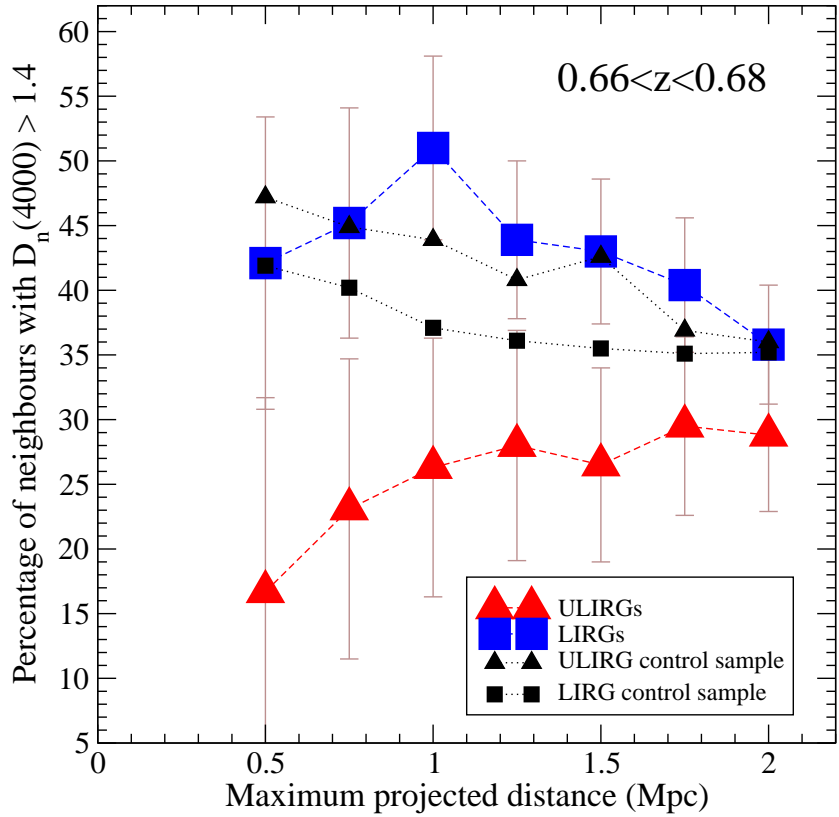}
\end{center}
\caption[]{\label{fig_dn4000dp} Percentages of (cumulative) neighbours with $\rm D_n(4000)> 1.4$ for the LIRGs, ULIRGs and their respective control samples, that lie on the density peak at $0.66<z<0.68$. All neighbours are selected within $\pm 500 \, \rm km \, s^{-1}$ of the central source. Symbols are the same as in figure \ref{fig_dn4000}.}
\end{figure}

We re-compute the percentages of (cumulative) close neighbours of \tfm galaxies which have $\rm D_n(4000)> 1.4$, versus the maximum projected physical distance, but this time considering only those IR galaxies at  $0.66<z<0.68$. The results are shown in Figure \ref{fig_dn4000dp}.

By comparing this figure with the left-hand panel of Figure \ref{fig_dn4000}  we clearly see that, within the density peak,  the percentages of close passive neighbours for LIRGs and optical galaxies in the control samples are higher  than the percentages obtained when considering the full redshift range $0.6<z<0.8$. This effect is consistent with the general fact that the fraction of red galaxies at $z\lsim1$ is larger in higher density environments (e.g. Hogg et al.~2003; Balogh et al.~2004; Baldry et al.~2006; Cooper et al.~2006;  Cucciati et al.~2006; Willmer et al.~2006). 

Surprisingly, instead, ULIRGs on the density peak are characterised by very active close environments, presenting no differences with respect to the total ULIRG population at $0.6<z<0.8$. It is particularly noteworthy the difference in the environment activities for ULIRGs and other optical galaxies with similar stellar masses. The high degree of activity in the ULIRG close neighbourhood suggests that this activity is directly linked to the ULIRG phase of the central galaxy, independently of the effects of structure at very large scale.

We note that the average number of close neighbours per ULIRG on the structure is higher than per LIRG ($\sim 5.4$ against $\sim 4$, within 2 Mpc of physical distance). However, the stellar mass ratio ($\rm M_{central}/M_{neighbour}$) distributions follow the same trends as those shown in the upper panel of Figure \ref{fig_mass}, i.e. LIRGs have companions of different relative stellar masses, while ULIRGs are virtually always the most massive galaxy of their close neighbourhood.

As a result, we find that the differential activities in the close environments of LIRGs and ULIRGs is especially enhanced in the density peak at $z\sim0.67$. Within 1 Mpc of projected physical distance, the 
percentages of passive neighbours are $\sim 51\%$ and $\sim 26\%$, for LIRGs and ULIRGs, respectively. Outside of the density peak, we find that there are milder but still non-negligible differences in the degrees of activity of LIRG and ULIRG neighbours at $0.6<z<0.8$ ($\sim$33\% against $\sim$23\% of passive neighbours within a distance of 1 Mpc, for LIRGs and ULIRGs, respectively).

%%%%%%%%%%%%%%%%%%%%%%%%%%%%%%%%%%%%%%%%%%%%%%%%%%%%%%%%%%%%%%%%%%%%%%%%%%%%%%%%%%%%%%%%%%%%
\section{\tfm galaxies in the \lowercase{z}COSMOS-reconstructed density fields}
\label{sec_densf}

The zCOSMOS-bright 10k catalogue has been used to reconstruct density fields up to $z=1$ (Kova\v{c} et al.~2008). It is of our interest to analyse the over-densities at the positions of the \tfm-selected sources in comparison to other optical galaxies with similar stellar masses.  A detailed explanation of the computation of these density fields is presented by Kova\v{c} et al.~(2008). Here, we consider only mass-weighted density fields computed using the 5th-nearest-neighbour method. These density fields are fully corrected for the sampling effects of the zCOSMOS-bright 10k sample.

\begin{figure*}
\begin{center}
\epsscale{1}
\plotone{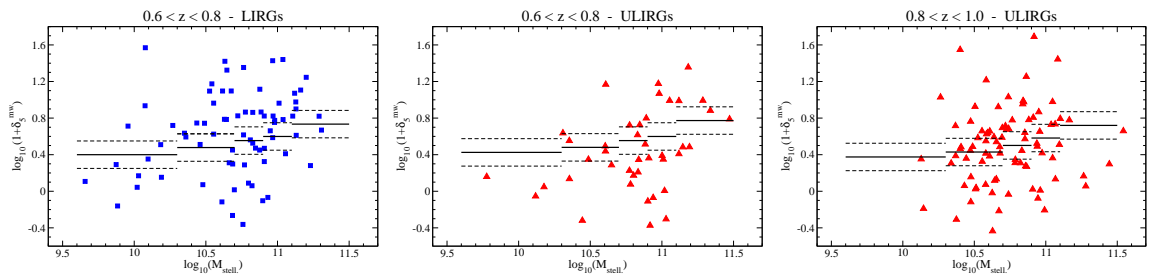}
\end{center}
\caption[]{\label{fig_df} Mass-weighted over-densities for LIRGs and ULIRGs, versus stellar masses (corresponding to a Chabrier IMF).  The horizontal lines indicate the mean and typical errors on the mass-weighted over-densities for the galaxies in the respective control samples, in different stellar mass bins.}
\end{figure*}

The three panels of Figure \ref{fig_df} show the mass-weighted over-densities  $(1+\delta_5^{\rm mw})$ for galaxies in our three IR galaxy populations, versus their stellar mass. For reference, we also plot the mean values and typical errors of the over-densities for galaxies in the respective control samples, in different stellar mass bins.  The typical errors on the  mass-weighted over-densities are  $\Delta \log_{10}(1+\delta_5^{\rm mw}) \lsim 0.15$, as determined with tests made on mock catalogues (Kova\v{c} et al.~2008).

We see that the IR galaxy over-densities are distributed around the mean values of the galaxies in the respective control samples. Globally, these results agree with the conclusions of  Marcillac et al.~(2008), who found that the differences between the environment densities of IR galaxies (LIRGs and ULIRGs altogether) and other optical galaxies at $0.7<z<1.0$ are not important, when both samples are restricted to have similar stellar masses.

However, within our sample at $0.6<z<0.8$, we find that there are twice as many ULIRGs in under-dense regions as in over-dense ones (i.e. beyond the error bars of the control sample means). Instead,  there are $\sim 70\%$ more LIRGs in over-dense regions than in under-dense ones. Finally, at $0.8<z<1.0$, the number of ULIRGs in our sample that are present in over- and under-dense regions are quite similar.  The differences in the general over-densities we find for LIRGs and ULIRGs are totally consistent with our results at smaller scales for the stellar-mass-ratio ($\rm M_{central}/M_{neighbour}$) distributions (see \S\ref{sec_envstm}).

Our findings are not in contradiction with Marcillac et al.~(2008) conclusions. If we considered our LIRGs and ULIRGs altogether, we would recover the result that IR sources at $0.6<z<1.0$ do not prefer either under- or over-dense environments with respect to other similarly massive galaxies.

%%%%%%%%%%%%%%%%%%%%%%%%%%%%%%%%%%%%%%%%%%%%%%%%%%%%%%%%%%%%%%%%%%%%%%%%%%%%%%%%%%%%%%%%%%
\section{Discussion}
\label{sec_concl}

 We conclude that LIRGs and ULIRGs at $0.6<z<0.8$ have close (within $2 \, \rm Mpc$ and $\pm \, 500 \, \rm km \, s^{-1}$) environments characterised by significantly different degrees of activity: the fraction of `passive' galaxies around LIRGs is substantially higher than the fraction around ULIRGs. Remarkably, the differential activity is especially enhanced in the COSMOS density field peak at $z\sim0.67$. By analysing control samples with the same stellar mass distributions as the two respective populations of IR galaxies, we show that the differences in environment activity are not an effect of the central galaxy stellar mass.

 Instead, this is probably a consequence of the IR-galaxy density fields. LIRGs and ULIRGs at $0.6<z<0.8$ are surrounded by close neighbours of different stellar masses, and their general local over-densities are also different. ULIRGs inhabit under-dense environments, while LIRGs  prefer over-denser regions. Thus, the differential activities observed in the  environments of the two populations are related to their density fields.
 
 At $0.8<z<1.0$, we also find that ULIRGs reside in more active close ($< 1 \, \rm Mpc$) environments than those of other optical galaxies with similar stellar masses. However, we do not find any difference between the stellar mass distributions of their close neighbours. This is in contrast with the environmental properties of ULIRGs at $0.6<z<0.8$. 
 
 At $0.8<z<1.0$, the similarities of the environments in which ULIRGs and other equally massive galaxies live suggest that these other galaxies might have also passed by a ULIRG phase at some moment in the past (consistently with the high incidence of ULIRGs among massive galaxies at $z \gsim 1.5$ found by Daddi et al.~2005 and Caputi et al.~2006b). The quenching of the IR phase in these galaxies could have happened along with the cessation of star formation in some of their close neighbours, raising the fraction of inactive sources in the surroundings.
 
 The number density of ULIRGs has a rapid decline at $z<1$ and, in particular, it reduces to around a half in the 1 Gyr of elapsed time between redshifts $z\sim0.9$ and $z\sim0.7$ (Le Floc'h et al.~2005; Caputi et al.~2007). This evolution occurs together with a change of environments, suggesting that the conditions under which ULIRGs occur are not the same at $z\sim0.9$ and $z\sim0.7$.  ULIRGs at $z\sim0.9$ are found in similar kinds of sites as other equally massive galaxies,  while those at $z\sim0.7$ appear mostly restricted to regions where they are the dominant galaxy of their close neighbourhood. It is plausible that the ULIRGs we observe at $z\sim0.7$ have been produced after the merging of close galaxies, leaving the final ULIRG close neighbourhood empty of massive neighbours. At $z\sim0.9$, even when mergers could also have been a possible mechanism for triggering the ULIRG phase, it seems that there has been no need to consume most of the close environment for these ULIRGs to be produced.

%____________________________________________
\acknowledgments

This paper is based on observations made with the VIMOS spectrograph on the VLT telescope, undertaken at the European Southern Observatory (ESO) under Large Program 175.A-0839. Also based on observations made with the {\em Spitzer} Observatory, which is operated under NASA contract 1407. 

%
%______________________________________________________________
% FINAL REFERENCES

\clearpage

%
% Redshift distribution
%______________________________________________________________

%
% Activity of close neighbours 
%______________________________________________________________

%
% Stellar mass ratios
%______________________________________________________________

%
% Activity of neighbours at z~0.67
%______________________________________________________________

% 
% Density fields
%______________________________________________________________ 

\end{document}